\documentclass[aps,prl,amsmath,lengthcheck,superscriptaddress,onecolumn]{revtex4-2}

\usepackage{graphicx}
\usepackage{hyperref}
\hypersetup{colorlinks,allcolors=blue,breaklinks}

\newcommand{\be}{\begin{equation}}
\newcommand{\ee}{\end{equation}}
\newcommand{\ba}{\begin{align}}
\newcommand{\ea}{\end{align}}
\newcommand{\bi}{\begin{itemize}}
\newcommand{\ei}{\end{itemize}}

\newcommand{\bla}{bla\\bla\\bla\\bla\\bla}

\newcommand{\mc}[1]{\mathcal{#1}}

\begin{document}

\title{Fluctuation theorem for time-averaged work}

\author{Pierre Naz\'e}
\email{pierre.naze@icen.ufpa.br}

\affiliation{\it Universidade Federal do Par\'a, Faculdade de F\'isica, ICEN,
Av. Augusto Correa, 1, Guam\'a, 66075-110, Bel\'em, Par\'a, Brazil}

\date{\today}

\begin{abstract}

There is evidence that taking the time average of the work performed by a thermally isolated system effectively "transforms" the adiabatic process into an isothermal one. This approach allows inherent quantities of adiabatic processes to be accessed through the definitions of isothermal processes. A fluctuation theorem is then established, linking the time-averaged work to the quasistatic work. Numerical evidence supporting this equality is provided for a classical harmonic oscillator with a driven linear equilibrium position parameter. Furthermore, the strong inequality for the average work is derived from the deduced fluctuation theorem using optimality arguments.

\end{abstract}

\maketitle

\section{Introduction}

Over the past decades, numerous fluctuation theorems have been established \cite{sevick2008,esposito2009,seifert2012,shiraishi2015fluctuation,crooks1999entropy,iyoda2017fluctuation,malek2017fluctuation,mandal2017entropy,pal2017integral,aaberg2018fully,lostaglio2018quantum,dabelow2019irreversibility,hasegawa2019fluctuation,micadei2020quantum,de2022quantum,bonancca2022fluctuation,mahault2022topological,murashita2022review,salazar2022thermodynamic,jarzynski1997}. These theorems establish a connection between non-equilibrium and equilibrium quantities, offering a fresh perspective on the Second Law of Thermodynamics, now articulated as an equality. The non-equilibrium quantity of interest is the work done on the system, which is described by the following random variable
\be
W({\bf z_0},\tau):=\int_0^\tau \partial_{\lambda_\tau(t)} \mathcal{H}({\bf z}({\bf z_0},t),\lambda_\tau(t))\dot{\lambda}_\tau(t)dt,
\ee
where $\partial_{\lambda_\tau(t)} \mathcal{H}({\bf z}({\bf z_0},t),\lambda_\tau(t))$ represents the generalized force associated with the system, ${\bf z}({\bf z}_0,t)$ is a point in the phase space $\Gamma$ of the system reached from the initial condition ${\bf z}_0$ in time $t$, and $\lambda_\tau(t)$ denotes the external protocol driving the system. Here, $\tau$ is the process duration of the driving performed. Historically, Jarzynski's work \cite{jarzynski1997} brought the fluctuation theorem into prominence, deriving from it the weak inequality
\be
\langle W\rangle(\tau)\ge\Delta F,
\label{eq:weakinq}
\ee
where $\langle W\rangle(\tau)$ represents the average work performed in a system at the initial conditions, and $\Delta F$ denotes the difference in Helmholtz free energy between the final and initial equilibrium states. The system in question can be either thermally isolated or open \cite{jarzynski1997}. In the first case, the system begins in a thermal state but evolves in time without interacting with the initial heat bath. In the second case, the system evolves while remaining in contact with the reservoir. This work focuses exclusively on thermally isolated systems. Additionally, the Helmholtz free energy at some parameter $\lambda$ is expressed as
\be
F(\lambda) = -\frac{1}{\beta}\ln{\left(\int_{\Gamma} e^{-\beta \mathcal{H}({\bf z},\lambda)}d{\bf z}\right)},
\label{eq:deltaf}
\ee
where $\beta:=(k_B T)^{-1}$, with $k_B$ being Boltzmann constant, and $T$ the temperature of the thermal bath. Its difference represents the trade-off between the energy and entropy of the system in an isothermal process performed at an infinitely slow time at the final and initial thermal equilibrium.

Another significant quantity for a thermally isolated system is the quasistatic work, $\langle W_{\rm qs}\rangle$, obtained when the average work in Eq.~\eqref{eq:weakinq} is varied infinitely slowly~\cite{acconcia2015,bonancca2016non,bonancca2019approaching,jarzynski2020,naze2022series}. The definition used in this work is
\be
\langle W_{\rm qs}\rangle:=\lim_{\tau\rightarrow \infty} \langle W\rangle(\tau).
\label{eq:wqsw}
\ee
In particular, the Second Law of Thermodynamics for this kind of system is a postulated stronger inequality
\be
\langle W\rangle(\tau) \ge \langle W_{\rm qs}\rangle,
\label{eq:strongine}
\ee
which has never been violated in any experiment~\cite{jarzynski2020}. This inequality is tighter than the weak one since $\langle W_{\rm qs}\rangle\ge \Delta F$, as dictated by the weak inequality for the quasistatic work. Notably, for an open system operating in an isothermal bath at an infinitely slow time, the quasistatic work of the entire system corresponds exactly to the difference of their Helmholtz free energies at the final and initial states, leading to the equality $\langle W_{\rm qs}\rangle = \Delta F$ in this scenario. Gaining insight into why such a strong inequality holds is important for advancing theoretical non-equilibrium statistical mechanics, particularly in shortcuts to adiabaticity~\cite{torrontegui2013shortcuts,guery2019shortcuts}.

Consider now the time-averaged work, given by the following random variable
\begin{equation}
    \overline{W}({\bf z_0},\tau) := \frac{1}{\tau}\int_0^\tau W({\bf z_0},t)dt.
    \label{eq:tawork}
\end{equation}
Previous studies have indicated that applying such time averaging to quantities defined for thermally isolated systems effectively ``transforms'' their adiabatic processes into isothermal processes of open systems~\cite{naze2023adiabatic,naze2023quantum}. This transformation enables the use of key definitions from isothermal processes, such as the relaxation time defined in linear response theory~\cite{naze2020compatibility}, in the context of adiabatic processes to identify analogous and inherent quantities. For example, a possible definition of a relaxation time in the phenomenon of the Kibble-Zurek mechanism was proposed in Ref.~\cite{naze2023quantum}. 

The time-averaged work becomes intuitive when compared to the work performed in an open system undergoing an isothermal process. In this case, to have the work depending only on the initial conditions of the system, a previous average is necessary to smooth down the influence of the noise. The same principle applies to thermally isolated systems performing an adiabatic process. Due to their inherent oscillatory qualities, like in their relaxation function in the realm of linear-response theory~\cite{naze2023adiabatic,naze2023quantum}, an average to smooth down such a behavior must occur. This average is exactly the time-averaged work. This idea could be extrapolated to other processes, like the isobaric ones, if the system exhibits some erratic behavior.

In this work, for thermally isolated systems undergoing an adiabatic driven process, I derive the following fluctuation theorem
\be
\langle e^{-\beta \overline{W}} \rangle = e^{-\beta \langle {W}_{\rm qs}\rangle},
\label{eq:jarzynskiequality}
\ee
Numerical evidence is presented for the classical harmonic oscillator with a driven linear equilibrium position parameter, expressed by 
\be
\mathcal{H}_1 = \frac{p^2}{2}+\frac{(q-\lambda_\tau(t))^2}{2},\quad \lambda_\tau(t)=\lambda_0+\frac{t}{\tau}\delta\lambda.
\label{eq:system}
\ee

Finally, I prove the strong inequality using optimality arguments and the weak inequality, $\langle \overline{W}\rangle(\tau)\ge \langle W_{\rm qs}\rangle$, derived from the deduced fluctuation theorem. In Jarzynski's work~\cite{jarzynski2020}, this inequality is proven for the ideal gas with a moving piston. In the end, the author states that the proof of the strong inequality for generic systems is still open.

\section{Preliminaries} 

Consider a system with Hamiltonian $\mathcal{H} [{\bf z}({\bf z}_0,t),\lambda_\tau(t)]$, dependent on some external parameter $\lambda_\tau(t)=\lambda_0+g(t/\tau)\delta\lambda$, with $g(0)=0$ and $g(1)=1$. Initially, the system is in thermal equilibrium, weakly coupled to a heat bath at a temperature $T$. When the process starts, the system is decoupled from the heat bath and adiabatically evolves in time, that is, without any source of heat. The work done on the system along the process is
\be
W({\bf z_0},\tau)  = \mathcal{H} [{\bf z}({\bf z_0},\tau),\lambda_0+\delta\lambda]-\mathcal{H} ({\bf z_0},\lambda_0),
\ee
or, as well,
\be
W({\bf z_0},\tau)  = \int_0^\tau \partial_\lambda \mathcal{H} [{\bf z}({\bf z_0},t),\lambda_\tau(t)]\dot{\lambda}_\tau(t)dt.
\ee
In its turn, the average work reads
\be
\langle W \rangle(\tau) = \int_\Gamma \mathcal{H} [{\bf z}({\bf z}_0,\tau),\lambda_0+\delta\lambda]\rho_{\mathcal{H}}[{\bf z}({\bf z}_0,\tau),\tau]d{\bf z}-\int_\Gamma \mathcal{H} ({\bf z_0},\lambda_0)\rho_{\mathcal{H}}({\bf z_0},0)d{\bf z_0},
\ee
or, as well,
\be
\langle W \rangle(\tau) = \int_\Gamma\int_0^\tau \partial_\lambda \mathcal{H} [{\bf z}({\bf z_0},t),\lambda_\tau(t)]\rho_{\mathcal{H}}[{\bf z}({\bf z_0},t),t]\dot{\lambda}_\tau(t)dtd{\bf z}.
\ee
The probability distribution $\rho_{\mathcal{H}}[{\bf z}({\bf z}_0,t),t]$ is the solution of the Liouville equation
\be
\frac{\partial\rho_{\mathcal{H}}}{\partial t} = \mathcal{L}_{\mathcal{H}}\rho_{\mathcal{H}},
\ee
where $\mathcal{L}_{\mathcal{H}}:=-\{\cdot,\mathcal{H}\}$ is the Liouville operator and $\rho_{\mc{H}}({\bf z_0},0)$ is the canonical ensemble. Here, $\{\cdot,\cdot\}$ is the Poisson bracket. 

Remembering Eq.~\eqref{eq:wqsw}, the so-called quasistatic work is the average work performed when the energy of the system is varied infinitely slowly
\be
\langle W_{\rm qs}\rangle:=\lim_{\tau\rightarrow\infty}\langle W\rangle(\tau)=\lim_{\tau\rightarrow\infty}\int_\Gamma\int_0^\tau \partial_\lambda \mathcal{H} [{\bf z}({\bf z}_0,t),\lambda_\tau(t)]\rho_{\mathcal{H}}[{\bf z}({\bf z}_0,t),t]\dot{\lambda}_\tau(t)dtd{\bf z}.
\label{eq:1stwqs}
\ee

\section{Time-averaged work} 

The quantity of interest here is the average of the time-averaged work, given by~\cite{naze2023adiabatic,naze2023quantum}
\be
\langle\overline{W}\rangle(\tau) = \frac{1}{\tau}\int_0^\tau \langle W\rangle(t)dt,
\ee
or, as well,
\be
\langle\overline{W}\rangle(\tau)  = \frac{1}{\tau}\int_0^\tau \left[\int_\Gamma \mathcal{H} [{\bf z}({\bf z}_0,t),\lambda_\tau(t)]\rho_{\mathcal{H}}[{\bf z}({\bf z}_0,t),t]d{\bf z}-\int_\Gamma \mathcal{H} ({\bf z_0},\lambda_0)\rho_{\mathcal{H}}({\bf z_0},0)d{\bf z_0}\right].
\label{eq:avwork}
\ee
Remark that such expression is the difference between two quantities. In this manner, rewriting in terms of a total derivative, one has
\be
\langle\overline{W}\rangle(\tau)  = \int_0^\tau \frac{d}{dt}\left[\frac{1}{t}\int_0^t\int_\Gamma \mathcal{H} [{\bf z}({\bf z}_0,t'),\lambda_t(t')]\rho_{\mathcal{H}}[{\bf z}({\bf z}_0,t'),t']d{\bf z}dt'\right]dt.
\ee
Note that taking the limit $t\rightarrow 0^+$ and using L'Hôspital rule, the following relation is true
\be
\lim_{t\rightarrow 0^+} \left\{\frac{1}{t}\int_0^t\int_\Gamma \mathcal{H} [{\bf z}({\bf z}_0,t'),\lambda_t(t')]\rho_{\mathcal{H}}[{\bf z}({\bf z}_0,t'),t']d{\bf z}dt'\right\} = \int_\Gamma \mathcal{H} ({\bf z_0},\lambda_0)\rho_{\mathcal{H}}({\bf z_0},0)d{\bf z_0}.
\ee
Observe that the average of the time-averaged work is an averaged work of the effective Hamiltonian
\begin{equation}
\mathcal{H}'[{\bf z'}({\bf z_0},t),t] := \frac{1}{t}\int_0^t \mathcal{H}\{{\bf z}[{\bf z'}({\bf z_0},t),t'],\lambda_t(t')\}dt',
\end{equation}
where ${\bf z'}({\bf z_0},t)$ is the solution of Hamilton's equations of the effective Hamiltonian. It is important to introduce such a function to define the pertinent quantities of the fluctuation theorem expressed in Eq.~\eqref{eq:jarzynskiequality}. Note also that $\mathcal{H}'[{\bf z'}({\bf z_0},0),0]=\mathcal{H}({\bf z_0},\lambda_0)$.

Consider now $\rho_{\mathcal{H}'}$ as the probability distribution associated with the effective Hamiltonian, where the initial probability distribution is equal to the canonical ensemble associated with $\mathcal{H}$. Using Liouville's theorem, one can show that
\be
\langle\overline{W}\rangle(\tau)  = \int_0^\tau\int_\Gamma \frac{d}{dt}\left[\frac{1}{t}\int_0^t\mathcal{H}\{{\bf z}[{\bf z'}({\bf z_0},t),t'],\lambda_t(t')\}dt'\right]\rho_{\mathcal{H}}({\bf z_0},0)d{\bf z_0}dt.
\ee
The random variable associated with the average of the time-averaged work reads
\be
\overline{W}({\bf z_0},\tau)  = \frac{1}{\tau}\int_0^\tau \mathcal{H}\{{\bf z}[{\bf z'}({\bf z_0},\tau),t],\lambda_\tau(t)\}dt-\mathcal{H} ({\bf z_0},\lambda_0),
\label{eq:tawork2}
\ee
confirming our initial definition expressed in Eq.~\eqref{eq:tawork}. 

To define the quasistatic time-averaged work $\langle\overline{W}_{\rm qs}\rangle$, apply Liouville's theorem in the first term of Eq.~\eqref{eq:avwork}. For infinitely slow processes, one has
\be
\langle\overline{W}_{\rm qs}\rangle=\lim_{\tau\rightarrow\infty}\int_\Gamma \left[\frac{1}{\tau}\int_0^\tau\mathcal{H} \{{\bf z}[{\bf z}'({\bf z_0},\tau),t],\lambda_\tau(t)\}dt\right]\rho_{\mathcal{H}}({\bf z_0},0)d{\bf z_0}-\int_\Gamma \mathcal{H}({\bf z_0},\lambda_0)\rho_{\mathcal{H}}({\bf z_0},0)d{\bf z_0}.
\label{eq:overwqs}
\ee
Remark that the quasistatic time-averaged work will be identical to the quasistatic work. Indeed,
\begin{align}
\langle\overline{W}_{\rm qs}\rangle &= \lim_{\tau\rightarrow\infty}\langle \overline{W}\rangle(\tau)\label{eq:TAwqs1}\\
&= \lim_{\tau\rightarrow\infty}\frac{1}{\tau}\int_0^\tau \langle W\rangle(t)dt\\
&= \int_0^1 \left[\lim_{\tau\rightarrow\infty}\langle W\rangle(\tau u)\right]du\\
&= \int_0^1 \langle W_{\rm qs} \rangle du\\
&=\langle W_{\rm qs} \rangle,
\label{eq:TAwqs2}
\end{align}
in which I used a change of variables in the integral in the second step. Remark the result is intuitive since the average of the time-averaged work smooths down the oscillatory behavior of the average work of the thermally isolated system, which is reduced to a constant value in infinitely slow processes.

Finally, the so-called difference in the time-averaged Helmholtz's free energies between the final and initial equilibrium states is defined. It is the difference of Helmholtz free energies of the effective Hamiltonian
\be
\Delta\overline{F}:=\overline{F}(\tau)-\overline{F}(0)=-\frac{1}{\beta}\ln\left(\frac{\int_{\Gamma}e^{-\beta \frac{1}{\tau}\int_0^\tau \mathcal{H} \{{\bf z}[{\bf z'({\bf z_0},\tau)},t],\lambda_\tau(t)\}dt}d{\bf z'_\tau}}{\int_{\Gamma}e^{-\beta \mathcal{H}({\bf z_0},\lambda_0)}d{\bf z_0}}\right).
\ee
Remark that such a definition is the difference between two quantities calculated at equilibrium. Its importance relies on its equivalence with the quasistatic work, discussed below.

\section{Equivalence between $\Delta \overline{F}$ and $W_{\rm qs}$}

To show the equivalence $\Delta \overline{F}=W_{\rm qs}$, consider
\be
\rho^c_{\mathcal{H}'}({\bf z}_t',t) := \frac{e^{-\beta \frac{1}{t}\int_0^t\mathcal{H} [{\bf z}({\bf z}_t',t'),\lambda_t(t')]dt'}}{\int_{\Gamma}e^{-\beta \frac{1}{t}\int_0^t\mathcal{H} [{\bf z}({\bf z}_t',t'),\lambda_t(t')]dt'}d{\bf z}_t'},
\label{eq:rhoc}
\ee
which is the probability distribution of the effective system and heat bath of temperature $T=(k_B \beta)^{-1}$ evolving infinitely slowly in time. The difference in the time-averaged Helmholtz's free energies can be rewritten as
\begin{align}
\Delta\overline{F} &=-\frac{1}{\beta}\ln\left\{\frac{\int_{\Gamma}e^{-\beta \frac{1}{\tau}\int_0^\tau \mathcal{H} [{\bf z}[({\bf z'_\tau},t),\lambda_\tau(t)]dt}d{\bf z'_\tau}}{\int_{\Gamma}e^{-\beta \mathcal{H}({\bf z_0},\lambda_0)}d{\bf z_0}}\right\}\\
&=-\frac{1}{\beta}\frac{d}{dt}\ln\left\{{\int_{\Gamma}e^{-\beta \frac{1}{t}\int_0^t\mathcal{H} [{\bf z}({\bf z}_t',t'),\lambda_t(t')]dt'}d{\bf z}_t'}\right\}\Bigg|_0^\tau\\
&=-\frac{1}{\beta}\left[\frac{\frac{d}{dt}\int_{\Gamma}e^{-\beta \frac{1}{t}\int_0^t\mathcal{H} [{\bf z}({\bf z}_t',t'),\lambda_t(t')]dt'}d{\bf z}_t'}{\int_{\Gamma}e^{-\beta \frac{1}{t}\int_0^t\mathcal{H} [{\bf z}({\bf z}_t',t'),\lambda_t(t')]dt'}d{\bf z}_t'}\right]\Bigg|_0^\tau\\
&= \lim_{\tau\rightarrow\infty}\int_0^\tau\int_\Gamma \frac{d}{dt}\left\{\frac{1}{t}\int_0^t\mathcal{H} [{\bf z}({\bf z}_t',t'),\lambda_t(t')]dt'\right\}\rho^c_{\mathcal{H}'}({\bf z}_t',t)d{\bf z}_t'dt.
\label{eq:overlinedeltaf1}
\end{align}
Since the process is adiabatic, the system dynamics is decoupled from the bath for all times considered in the time-integral within the derivative in Eq.~\eqref{eq:overlinedeltaf1}. However, there is an exception for the initial instant, which depends only on the Hamiltonian of the system in the average with the probability distribution of Eq.~\eqref{eq:rhoc}. In this manner, Eq.~\eqref{eq:overlinedeltaf1}, when calculated at the final and initial points, furnishes the average difference of energy of the thermally isolated system. This leads to
\be
\Delta\overline{F}= \lim_{\tau\rightarrow\infty}\int_\Gamma \left[\frac{1}{\tau}\int_0^\tau\mathcal{H} [{\bf z}({\bf z}_\tau',t),\lambda_\tau(t)]dt\right]\rho^c_{\mathcal{H}'}({\bf z}_\tau',t)d{\bf z}_\tau'-\int_{\Gamma}\mathcal{H}({\bf z_0},\lambda_0)\rho_{\mathcal{H}}({\bf z_0},0)d{\bf z_0}.
\ee
Using Liouville's theorem in the first term
\be
\Delta\overline{F} = \lim_{\tau\rightarrow\infty}\int_\Gamma \left[\frac{1}{\tau}\int_0^\tau\mathcal{H} \{{\bf z}[{\bf z}'({\bf z_0},\tau),t],\lambda_\tau(t)\}dt\right]\rho_{\mathcal{H}}({\bf z_0},0)d{\bf z_0}-\int_\Gamma \mathcal{H}({\bf z_0},\lambda_0)\rho_{\mathcal{H}}({\bf z_0},0)d{\bf z_0}.
\label{eq:overdeltaf}
\ee
Then the variation of the time-averaged Helmholtz's free energies between the final and initial state is nothing more than the work of the effective system performed in a quasistatic process. Therefore, 
\begin{equation}
    \Delta\overline{F}=\langle\overline{W}_{\rm qs}\rangle=\langle{W}_{\rm qs}\rangle,
\end{equation}
where in the first equality I compare Eq.~\eqref{eq:overdeltaf} with Eq.~\eqref{eq:overwqs}, while in the second one the relation deduced in Eqs.~\eqref{eq:TAwqs1}-\eqref{eq:TAwqs2} was used. In this manner, the time-averaged work ``cleans" the intrinsic oscillatory behavior of thermally isolated systems without being coupled to a heat bath, whose noise behavior is smoothed down in an average. Also, this illustrates that an adiabatic process quantity was obtained with the time-averaged work via a definition of an isothermal process.

Remark that the quasistatic work has a more tractable expression, given by
\be
\langle W_{\rm qs}\rangle=-\frac{1}{\beta}\ln\left(\frac{\int_{\Gamma}e^{-\beta \frac{1}{\tau}\int_0^\tau \mathcal{H} ({\bf z}({\bf z'_\tau},t),\lambda_\tau(t))dt}d{\bf z'_\tau}}{\int_{\Gamma}e^{-\beta \mathcal{H}({\bf z_0},\lambda_0)}d{\bf z_0}}\right),
\label{eq:newWqs}
\ee
than using the definition with the adiabatic invariant~\cite{fasano2006analytical}. To see whether the result is reasonable, I shall analyze the difference $\langle W_{\rm qs}\rangle-\Delta F$ under linear response theory. This is done considering an expansion of the relative change of the control parameter $\delta\lambda/\lambda_0$, with $\delta\lambda/\lambda_0\ll 1$. Up to second-order, the result is approximately
\begin{equation}
    \langle W_{\rm qs}\rangle-\Delta F\approx \frac{\beta}{2}(\langle \partial_{\lambda_0} \mathcal{H}({\bf z}_0,\lambda_0)^2\rangle-\langle \partial_{\lambda_0} \mathcal{H}({\bf z}_0,\lambda_0)\rangle^2)\delta\lambda^2,
\label{eq:wqsminusdeltaf}
\end{equation}
which is a positive number and converges to zero for open systems whose work probability distribution goes to a Dirac delta~\cite{jarzynski1997}. Consider now the result derived with the definition using adiabatic invariant of $\langle W_{\rm qs}\rangle$ in Ref.~\cite{acconcia2015}, for the driven stiffening parameter harmonic oscillator, given by
\be
\mathcal{H}_2 = \frac{p^2}{2}+\lambda_\tau(t)\frac{q^2}{2}.
\ee
Calculating its $\Delta F$ using Eq.~\eqref{eq:deltaf}, and comparing with Eq.~\eqref{eq:wqsminusdeltaf}, both results match, with
\begin{equation}
    \langle W_{\rm qs}\rangle-\Delta F\approx \frac{\delta\lambda^2}{8\beta\lambda_0^2}.
\end{equation}
This result corroborates the new expression of Eq.~\eqref{eq:newWqs}.

\section{Fluctuation theorem for time-averaged work}

\begin{figure}[h]
    \centering
    \includegraphics[scale=0.5]{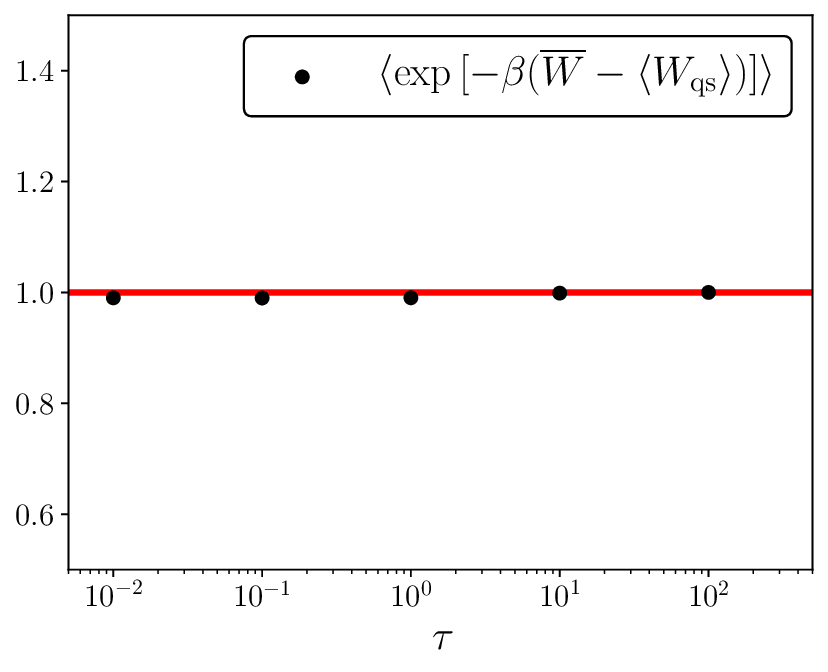}
    \caption{Fluctuation theorem~\eqref{eq:jarzynskiequality} for the classical harmonic oscillator, with a driven linear equilibrium position parameter. It was used datasets with $10^5$ values of $\exp{[-\beta(\overline{W}-\langle{W}_{\rm qs}\rangle)]}$ sampled according with the canonical ensemble. It was used $\beta=1$, $\lambda_0=1$, $\delta\lambda=0.5$.}
    \label{fig:JE}
\end{figure}

Now I derive the fluctuation theorem associated with the time-averaged work. Following the idea used in Ref.~\cite{jarzynski1997}, note
\begin{align}
\langle e^{-\beta \overline{W}} \rangle &= \int_{\Gamma} \exp\left(-\beta\left(\frac{1}{\tau}\int_0^\tau \mathcal{H}({\bf z}({\bf z'_\tau},t),\lambda_\tau(t))dt\right)+\beta\mathcal{H}({\bf z_0},\lambda_0)\right)\rho_{\mathcal{H}}({\bf z_0},0)d{\bf z'_\tau}\\
&= \int_{\Gamma} \exp\left(-\beta\left(\frac{1}{\tau}\int_0^\tau \mathcal{H}({\bf z}({\bf z'_\tau},t),\lambda_\tau(t))dt\right)\right)\exp(\beta\mathcal{H}({\bf z_0},\lambda_0))\frac{\exp(-\beta\mathcal{H}({\bf z_0},\lambda_0))}{\int_\Gamma \exp(\beta\mathcal{H}({\bf z_0},\lambda_0))d{\bf z_0}}d{\bf z'_\tau}\\
&=\frac{\int_{\Gamma} \exp\left(-\beta\left(\frac{1}{\tau}\int_0^\tau \mathcal{H}({\bf z}({\bf z'_\tau},t),\lambda_\tau(t))dt\right)d{\bf z'_\tau}\right)}{\int_{\Gamma} \exp\left(-\beta\mathcal{H}({\bf z_0},\lambda_0)\right)d{\bf z_0}}\\
&=e^{-\beta \Delta\overline{F}}\\
&=e^{-\beta \langle\overline{W}_{\rm qs}\rangle}\\
&=e^{-\beta \langle W_{\rm qs}\rangle}.
\label{eq:JE1}
\end{align}
Therefore, the fluctuation theorem expressed in Eq.~\eqref{eq:jarzynskiequality} holds. To corroborate this equality, I test the classical harmonic oscillator with a driven linear equilibrium position parameter, expressed in Eq.~\eqref{eq:system}. Figure~\ref{fig:JE} presents the result of the simulation. For each process duration $\tau$ I used datasets with $N=10^5$ values of $\exp{[-\beta(\overline{W}-\langle{W}_{\rm qs}\rangle)]}$, sampled accordingly with the canonical ensemble, and parameters $\beta=1$, $\lambda_0=1$ and $\delta\lambda=0.5$. The results for rapid protocols deviate about $1\%$ from the expected result. I justify those errors due to the inherent difficulty of sampling exponentially weighted random variables, primarily in this type of process whose work probability distributions are wider and require the occurrence of rare events \cite{jarzynski2006}.  

To finalize such section, I prove the weak inequality for the fluctuation theorem of the time-averaged work. Considering Jensen's inequality, one has
\be
e^{-\beta \langle {W}_{\rm qs}\rangle}=\langle e^{-\beta \overline{W}} \rangle \geq e^{-\beta \langle \overline{W}\rangle}.
\label{eq:inequalityft}
\ee
Using the increasing property of the exponential function and $-\beta<0$, we finally have 
\begin{equation}
    \langle \overline{W}\rangle(\tau)\ge \langle W_{\rm qs}\rangle.
    \label{eq:weakinq2}
\end{equation}

\section{Proving the strong inequality}

To prove the strong inequality, $\langle W\rangle(\tau)\ge \langle W_{\rm qs}\rangle$, one needs to verify that $\langle W_{\rm qs}\rangle$ represents a lower bound for all process durations and protocols used. The strategy is to demonstrate first that the optimal average work is equal to the optimal average of the time-averaged work. Then, by applying the optimality condition and the weak inequality~\eqref{eq:weakinq2} from the deduced fluctuation theorem, the inequality is proven.

Consider a thermally isolated system where the work is calculated using a fixed protocol for an external parameter. Let us find its local extremum. According to the definition of time-averaged work, one has
\be
\langle \overline{W}\rangle(\tau)=\frac{1}{\tau}\int_0^\tau \langle W\rangle(t)dt.
\label{eq:def2}
\ee
To treat the expression of the average of the time-averaged and averaged work in a more tractable way, let us take the derivative in $\tau$ of Eq.~\eqref{eq:def2}
\begin{align}
\frac{d}{d\tau}\left[\tau\langle \overline{W}\rangle(\tau)\right]&=\frac{d}{d\tau}\left[\int_0^\tau \langle W\rangle(t)dt\right]\\
\langle\overline{W}\rangle(\tau)+\tau\langle\dot{\overline{W}}\rangle(\tau)&= \langle W\rangle(\tau).
\label{eq:1stderivative}
\end{align}
Choosing $\tau^*$ where $\langle\dot{\overline{W}}\rangle(\tau^*)$ is equal to zero, one has 
\begin{equation}
\langle\overline{W}\rangle^*:=\langle\overline{W}\rangle(\tau^*)= \langle W\rangle(\tau^*),
\end{equation}
meaning that the instants where $\langle\overline{W}\rangle^*$ is an extremum, it is equal to $\langle W\rangle(\tau^*)$. To show the case where $\langle W\rangle^*:=\langle W\rangle(\tau^*)$ is an extremum, and being equal to $\langle \overline{W}\rangle^*$, consider the derivative of Eq.~\eqref{eq:1stderivative} in $\tau$
\begin{align}
\frac{d}{d\tau}[\langle\overline{W}\rangle(\tau)+\tau\langle\dot{\overline{W}}\rangle(\tau)] &= \frac{d}{d\tau}[\langle W\rangle(\tau)]\\
2\langle\dot{\overline{W}}\rangle(\tau)+\tau\langle\ddot{\overline{W}}\rangle(\tau)&= \langle \dot{W}\rangle(\tau).
\label{eq:2ndderivative}
\end{align}
Assuming that exists a $\tau^*$ where $\langle W\rangle(\tau^*)$ is an extremum, observe that 
\be
2\langle\dot{\overline{W}}\rangle(\tau^*)=-\tau^*\langle\ddot{\overline{W}}\rangle(\tau^*).
\label{eq:2ndder2}
\ee
Note also
\begin{align}
    \tau\langle\ddot{\overline{W}}\rangle(\tau)&=\tau \frac{d^2}{d\tau^2}\left[ \frac{1}{\tau}\int_0^\tau \langle W\rangle(t) dt\right]\\
    &=\tau \frac{d}{d\tau}\left[ \frac{\langle\overline{W}\rangle(\tau)-\langle W\rangle(\tau)}{\tau}\right]\\
    &= \frac{\langle\overline{W}\rangle(\tau)-\langle W\rangle(\tau)}{\tau}-(\langle\dot{\overline{W}}\rangle(\tau)-\langle \dot{W}\rangle(\tau))\\
    &= \langle\dot{W}\rangle(\tau).
\label{eq:id}
\end{align}
Plugging Eq.~\eqref{eq:id} in Eq.~\eqref{eq:2ndder2}, this will imply that $\langle \overline{W}\rangle(\tau^*)$ is an extremum. Therefore, it holds 
\begin{equation}
\langle\overline{W}\rangle^*= \langle W\rangle^*,
\end{equation}
showing that the local extremum of the time-averaged work is equal to the averaged work one. However, this relation holds locally for this type of optimization process. Let us now proceed with global optimization and observe that the strong inequality holds. Note that the global maximum and minimum occur either at the boundaries of the interval or at local maxima and minima within it. Therefore, there are three possible options for the global minimum:

\vspace{0.25cm}

1. $\tau^*\rightarrow 0^+$

\vspace{0.25cm}

In this limit, the initial Hamiltonian abruptly changes the external parameter, while the probability distribution does not. The same happens to the effective Hamiltonian. Therefore
\begin{equation}
    \langle W\rangle^*=\langle W\rangle(0^+)=\langle \mathcal{H}(\lambda_0+\delta\lambda)\rangle-\langle \mathcal{H}(\lambda_0)\rangle=\langle \overline{W}\rangle(0^+)=\langle \overline{W}\rangle^*.
\end{equation}
Using the optimality condition and the weak inequality~\eqref{eq:weakinq2}, 
one has
\begin{equation}
    \langle W\rangle(\tau) \ge \langle W\rangle^*=\langle \overline{W}\rangle^*\ge \langle W_{\rm qs}\rangle.
\end{equation}

\vspace{0.25cm}

2. $\tau^*\rightarrow \infty$

\vspace{0.25cm}

This case is straightforward
\begin{equation}
    \langle W\rangle(\tau) \ge\langle W\rangle^*=\lim_{\tau\rightarrow\infty}\langle W\rangle(\tau)=\langle W_{\rm qs}\rangle.
\end{equation}

\vspace{0.25cm}

3. $0<\tau^*<\infty$

\vspace{0.25cm}

Using the results deduced for local extremum, one has
\begin{equation}
    \langle W\rangle(\tau)\ge \langle W\rangle^*=\langle \overline{W}\rangle^*\ge\langle W_{\rm qs}\rangle.
\end{equation}
Note that no inequality can be proposed about the global maximum. Therefore, a strong inequality exists for that particular protocol of the external parameter used to calculate the work at the beginning of this section. However, since the quasistatic work does not depend on the protocol, the strong inequality holds for all $\tau$ and all protocols used. Consequently, the strategy succeeded primarily due to the equality between the optimal average works calculated at $\tau^*$.

\section{Conclusion} 

In the present work, a fluctuation theorem is established, linking the time-averaged work to the quasistatic work for thermally isolated systems undergoing an adiabatic driving process. It is assumed here that the system is classical and starts weakly coupled to a heat bath, which is removed when the driving begins. Numerical evidence supporting this fluctuation theorem is provided for the classical harmonic oscillator with a driven linear equilibrium position parameter. The findings confirm that the time-averaged work "transforms" a thermally isolated system performing an adiabatic process into an open system performing an isothermal one, where the inherent oscillatory behavior is smoothed down, like in the relaxation function in the realm of linear response theory~\cite{naze2023adiabatic,naze2023quantum}. The strong inequality, which has never been violated in any experiment, is derived from first principles using the deduced fluctuation theorem and optimality arguments. This result suggests that no protocols better than shortcuts to adiabaticity~\cite{torrontegui2013shortcuts,guery2019shortcuts} can exist under the conditions outlined in this work. However, there is evidence that the strong inequality breaks down for nonlinear systems starting in the microcanonical ensemble~\cite{acconcia2017microcanonical}. This will be a topic to be explored in future research.

\newpage

\bibliography{FTTA.bib}

\begin{thebibliography}{33}%
\makeatletter
\providecommand \@ifxundefined [1]{%
 \@ifx{#1\undefined}
}%
\providecommand \@ifnum [1]{%
 \ifnum #1\expandafter \@firstoftwo
 \else \expandafter \@secondoftwo
 \fi
}%
\providecommand \@ifx [1]{%
 \ifx #1\expandafter \@firstoftwo
 \else \expandafter \@secondoftwo
 \fi
}%
\providecommand \natexlab [1]{#1}%
\providecommand \enquote  [1]{``#1''}%
\providecommand \bibnamefont  [1]{#1}%
\providecommand \bibfnamefont [1]{#1}%
\providecommand \citenamefont [1]{#1}%
\providecommand \href@noop [0]{\@secondoftwo}%
\providecommand \href [0]{\begingroup \@sanitize@url \@href}%
\providecommand \@href[1]{\@@startlink{#1}\@@href}%
\providecommand \@@href[1]{\endgroup#1\@@endlink}%
\providecommand \@sanitize@url [0]{\catcode `\\12\catcode `\$12\catcode
  `\&12\catcode `\#12\catcode `\^12\catcode `\_12\catcode `\%12\relax}%
\providecommand \@@startlink[1]{}%
\providecommand \@@endlink[0]{}%
\providecommand \url  [0]{\begingroup\@sanitize@url \@url }%
\providecommand \@url [1]{\endgroup\@href {#1}{\urlprefix }}%
\providecommand \urlprefix  [0]{URL }%
\providecommand \Eprint [0]{\href }%
\providecommand \doibase [0]{https://doi.org/}%
\providecommand \selectlanguage [0]{\@gobble}%
\providecommand \bibinfo  [0]{\@secondoftwo}%
\providecommand \bibfield  [0]{\@secondoftwo}%
\providecommand \translation [1]{[#1]}%
\providecommand \BibitemOpen [0]{}%
\providecommand \bibitemStop [0]{}%
\providecommand \bibitemNoStop [0]{.\EOS\space}%
\providecommand \EOS [0]{\spacefactor3000\relax}%
\providecommand \BibitemShut  [1]{\csname bibitem#1\endcsname}%
\let\auto@bib@innerbib\@empty
\bibitem [{\citenamefont {Sevick}\ \emph {et~al.}(2008)\citenamefont {Sevick},
  \citenamefont {Prabhakar}, \citenamefont {Williams},\ and\ \citenamefont
  {Searles}}]{sevick2008}%
  \BibitemOpen
  \bibfield  {author} {\bibinfo {author} {\bibfnamefont {E.~M.}\ \bibnamefont
  {Sevick}}, \bibinfo {author} {\bibfnamefont {R.}~\bibnamefont {Prabhakar}},
  \bibinfo {author} {\bibfnamefont {S.~R.}\ \bibnamefont {Williams}},\ and\
  \bibinfo {author} {\bibfnamefont {D.~J.}\ \bibnamefont {Searles}},\ }\href
  {https://www.annualreviews.org/doi/abs/10.1146/annurev.physchem.58.032806.104555?casa_token=hHvU7UHFgUcAAAAA\%3ALaJNzI8pK_gnKIS_WplRga55tOPflCjoAXynUitOB7Sl8rT_1GAd3oRkZ9mPr8WpYeJLe1mdeoh4mJX8}
  {\bibfield  {journal} {\bibinfo  {journal} {Annu. Rev. Phys. Chem.}\ }\textbf
  {\bibinfo {volume} {59}},\ \bibinfo {pages} {603} (\bibinfo {year}
  {2008})}\BibitemShut {NoStop}%
\bibitem [{\citenamefont {Esposito}\ \emph {et~al.}(2009)\citenamefont
  {Esposito}, \citenamefont {Harbola},\ and\ \citenamefont
  {Mukamel}}]{esposito2009}%
  \BibitemOpen
  \bibfield  {author} {\bibinfo {author} {\bibfnamefont {M.}~\bibnamefont
  {Esposito}}, \bibinfo {author} {\bibfnamefont {U.}~\bibnamefont {Harbola}},\
  and\ \bibinfo {author} {\bibfnamefont {S.}~\bibnamefont {Mukamel}},\ }\href
  {https://journals.aps.org/rmp/abstract/10.1103/RevModPhys.81.1665} {\bibfield
   {journal} {\bibinfo  {journal} {Reviews of modern physics}\ }\textbf
  {\bibinfo {volume} {81}},\ \bibinfo {pages} {1665} (\bibinfo {year}
  {2009})}\BibitemShut {NoStop}%
\bibitem [{\citenamefont {Seifert}(2012)}]{seifert2012}%
  \BibitemOpen
  \bibfield  {author} {\bibinfo {author} {\bibfnamefont {U.}~\bibnamefont
  {Seifert}},\ }\href
  {https://iopscience.iop.org/article/10.1088/0034-4885/75/12/126001/meta?casa_token=Ua1rySVLERMAAAAA:yhdj5HMg32DWqQ2z3lDVIWjD2h6OV_P-6q56bDnG7nUR6tN5tNDkK2vmH9EL3cqYWdt1e1u1b8A}
  {\bibfield  {journal} {\bibinfo  {journal} {Reports on progress in physics}\
  }\textbf {\bibinfo {volume} {75}},\ \bibinfo {pages} {126001} (\bibinfo
  {year} {2012})}\BibitemShut {NoStop}%
\bibitem [{\citenamefont {Shiraishi}\ and\ \citenamefont
  {Sagawa}(2015)}]{shiraishi2015fluctuation}%
  \BibitemOpen
  \bibfield  {author} {\bibinfo {author} {\bibfnamefont {N.}~\bibnamefont
  {Shiraishi}}\ and\ \bibinfo {author} {\bibfnamefont {T.}~\bibnamefont
  {Sagawa}},\ }\href {https://doi.org/10.1103/PhysRevE.91.012130} {\bibfield
  {journal} {\bibinfo  {journal} {Physical Review E}\ }\textbf {\bibinfo
  {volume} {91}},\ \bibinfo {pages} {012130} (\bibinfo {year}
  {2015})}\BibitemShut {NoStop}%
\bibitem [{\citenamefont {Crooks}(1999)}]{crooks1999entropy}%
  \BibitemOpen
  \bibfield  {author} {\bibinfo {author} {\bibfnamefont {G.~E.}\ \bibnamefont
  {Crooks}},\ }\href {https://doi.org/10.1103/PhysRevE.60.2721} {\bibfield
  {journal} {\bibinfo  {journal} {Physical Review E}\ }\textbf {\bibinfo
  {volume} {60}},\ \bibinfo {pages} {2721} (\bibinfo {year}
  {1999})}\BibitemShut {NoStop}%
\bibitem [{\citenamefont {Iyoda}\ \emph {et~al.}(2017)\citenamefont {Iyoda},
  \citenamefont {Kaneko},\ and\ \citenamefont {Sagawa}}]{iyoda2017fluctuation}%
  \BibitemOpen
  \bibfield  {author} {\bibinfo {author} {\bibfnamefont {E.}~\bibnamefont
  {Iyoda}}, \bibinfo {author} {\bibfnamefont {K.}~\bibnamefont {Kaneko}},\ and\
  \bibinfo {author} {\bibfnamefont {T.}~\bibnamefont {Sagawa}},\ }\href
  {https://doi.org/10.1103/PhysRevLett.119.100601} {\bibfield  {journal}
  {\bibinfo  {journal} {Physical review letters}\ }\textbf {\bibinfo {volume}
  {119}},\ \bibinfo {pages} {100601} (\bibinfo {year} {2017})}\BibitemShut
  {NoStop}%
\bibitem [{\citenamefont {Malek~Mansour}\ and\ \citenamefont
  {Baras}(2017)}]{malek2017fluctuation}%
  \BibitemOpen
  \bibfield  {author} {\bibinfo {author} {\bibfnamefont {M.}~\bibnamefont
  {Malek~Mansour}}\ and\ \bibinfo {author} {\bibfnamefont {F.}~\bibnamefont
  {Baras}},\ }\href {https://doi.org/10.1063/1.4986600} {\bibfield  {journal}
  {\bibinfo  {journal} {Chaos: An Interdisciplinary Journal of Nonlinear
  Science}\ }\textbf {\bibinfo {volume} {27}} (\bibinfo {year}
  {2017})}\BibitemShut {NoStop}%
\bibitem [{\citenamefont {Mandal}\ \emph {et~al.}(2017)\citenamefont {Mandal},
  \citenamefont {Klymko},\ and\ \citenamefont {DeWeese}}]{mandal2017entropy}%
  \BibitemOpen
  \bibfield  {author} {\bibinfo {author} {\bibfnamefont {D.}~\bibnamefont
  {Mandal}}, \bibinfo {author} {\bibfnamefont {K.}~\bibnamefont {Klymko}},\
  and\ \bibinfo {author} {\bibfnamefont {M.~R.}\ \bibnamefont {DeWeese}},\
  }\href {https://doi.org/10.1103/PhysRevLett.119.258001} {\bibfield  {journal}
  {\bibinfo  {journal} {Physical review letters}\ }\textbf {\bibinfo {volume}
  {119}},\ \bibinfo {pages} {258001} (\bibinfo {year} {2017})}\BibitemShut
  {NoStop}%
\bibitem [{\citenamefont {Pal}\ and\ \citenamefont
  {Rahav}(2017)}]{pal2017integral}%
  \BibitemOpen
  \bibfield  {author} {\bibinfo {author} {\bibfnamefont {A.}~\bibnamefont
  {Pal}}\ and\ \bibinfo {author} {\bibfnamefont {S.}~\bibnamefont {Rahav}},\
  }\href {https://doi.org/10.1103/PhysRevE.96.062135} {\bibfield  {journal}
  {\bibinfo  {journal} {Physical Review E}\ }\textbf {\bibinfo {volume} {96}},\
  \bibinfo {pages} {062135} (\bibinfo {year} {2017})}\BibitemShut {NoStop}%
\bibitem [{\citenamefont {{\AA}berg}(2018)}]{aaberg2018fully}%
  \BibitemOpen
  \bibfield  {author} {\bibinfo {author} {\bibfnamefont {J.}~\bibnamefont
  {{\AA}berg}},\ }\href {Fully quantum fluctuation theorems} {\bibfield
  {journal} {\bibinfo  {journal} {Physical Review X}\ }\textbf {\bibinfo
  {volume} {8}},\ \bibinfo {pages} {011019} (\bibinfo {year}
  {2018})}\BibitemShut {NoStop}%
\bibitem [{\citenamefont {Lostaglio}(2018)}]{lostaglio2018quantum}%
  \BibitemOpen
  \bibfield  {author} {\bibinfo {author} {\bibfnamefont {M.}~\bibnamefont
  {Lostaglio}},\ }\href {https://doi.org/10.1103/PhysRevLett.120.040602}
  {\bibfield  {journal} {\bibinfo  {journal} {Physical review letters}\
  }\textbf {\bibinfo {volume} {120}},\ \bibinfo {pages} {040602} (\bibinfo
  {year} {2018})}\BibitemShut {NoStop}%
\bibitem [{\citenamefont {Dabelow}\ \emph {et~al.}(2019)\citenamefont
  {Dabelow}, \citenamefont {Bo},\ and\ \citenamefont
  {Eichhorn}}]{dabelow2019irreversibility}%
  \BibitemOpen
  \bibfield  {author} {\bibinfo {author} {\bibfnamefont {L.}~\bibnamefont
  {Dabelow}}, \bibinfo {author} {\bibfnamefont {S.}~\bibnamefont {Bo}},\ and\
  \bibinfo {author} {\bibfnamefont {R.}~\bibnamefont {Eichhorn}},\ }\href
  {https://doi.org/10.1103/PhysRevX.9.021009} {\bibfield  {journal} {\bibinfo
  {journal} {Physical Review X}\ }\textbf {\bibinfo {volume} {9}},\ \bibinfo
  {pages} {021009} (\bibinfo {year} {2019})}\BibitemShut {NoStop}%
\bibitem [{\citenamefont {Hasegawa}\ and\ \citenamefont
  {Van~Vu}(2019)}]{hasegawa2019fluctuation}%
  \BibitemOpen
  \bibfield  {author} {\bibinfo {author} {\bibfnamefont {Y.}~\bibnamefont
  {Hasegawa}}\ and\ \bibinfo {author} {\bibfnamefont {T.}~\bibnamefont
  {Van~Vu}},\ }\href {https://doi.org/10.1103/PhysRevLett.123.110602}
  {\bibfield  {journal} {\bibinfo  {journal} {Physical review letters}\
  }\textbf {\bibinfo {volume} {123}},\ \bibinfo {pages} {110602} (\bibinfo
  {year} {2019})}\BibitemShut {NoStop}%
\bibitem [{\citenamefont {Micadei}\ \emph {et~al.}(2020)\citenamefont
  {Micadei}, \citenamefont {Landi},\ and\ \citenamefont
  {Lutz}}]{micadei2020quantum}%
  \BibitemOpen
  \bibfield  {author} {\bibinfo {author} {\bibfnamefont {K.}~\bibnamefont
  {Micadei}}, \bibinfo {author} {\bibfnamefont {G.~T.}\ \bibnamefont {Landi}},\
  and\ \bibinfo {author} {\bibfnamefont {E.}~\bibnamefont {Lutz}},\ }\href
  {https://doi.org/10.1103/PhysRevLett.124.090602} {\bibfield  {journal}
  {\bibinfo  {journal} {Physical Review Letters}\ }\textbf {\bibinfo {volume}
  {124}},\ \bibinfo {pages} {090602} (\bibinfo {year} {2020})}\BibitemShut
  {NoStop}%
\bibitem [{\citenamefont {De~Chiara}\ and\ \citenamefont
  {Imparato}(2022)}]{de2022quantum}%
  \BibitemOpen
  \bibfield  {author} {\bibinfo {author} {\bibfnamefont {G.}~\bibnamefont
  {De~Chiara}}\ and\ \bibinfo {author} {\bibfnamefont {A.}~\bibnamefont
  {Imparato}},\ }\href {https://doi.org/10.1103/PhysRevResearch.4.023230}
  {\bibfield  {journal} {\bibinfo  {journal} {Physical Review Research}\
  }\textbf {\bibinfo {volume} {4}},\ \bibinfo {pages} {023230} (\bibinfo {year}
  {2022})}\BibitemShut {NoStop}%
\bibitem [{\citenamefont {Bonan{\c{c}}a}\ and\ \citenamefont
  {Deffner}(2022)}]{bonancca2022fluctuation}%
  \BibitemOpen
  \bibfield  {author} {\bibinfo {author} {\bibfnamefont {M.~V.~S.}\
  \bibnamefont {Bonan{\c{c}}a}}\ and\ \bibinfo {author} {\bibfnamefont
  {S.}~\bibnamefont {Deffner}},\ }\href
  {https://doi.org/10.1103/PhysRevE.105.L012105} {\bibfield  {journal}
  {\bibinfo  {journal} {Physical Review E}\ }\textbf {\bibinfo {volume}
  {105}},\ \bibinfo {pages} {L012105} (\bibinfo {year} {2022})}\BibitemShut
  {NoStop}%
\bibitem [{\citenamefont {Mahault}\ \emph {et~al.}(2022)\citenamefont
  {Mahault}, \citenamefont {Tang},\ and\ \citenamefont
  {Golestanian}}]{mahault2022topological}%
  \BibitemOpen
  \bibfield  {author} {\bibinfo {author} {\bibfnamefont {B.}~\bibnamefont
  {Mahault}}, \bibinfo {author} {\bibfnamefont {E.}~\bibnamefont {Tang}},\ and\
  \bibinfo {author} {\bibfnamefont {R.}~\bibnamefont {Golestanian}},\ }\href
  {https://doi.org/10.1038/s41467-022-30644-6} {\bibfield  {journal} {\bibinfo
  {journal} {Nature Communications}\ }\textbf {\bibinfo {volume} {13}},\
  \bibinfo {pages} {3036} (\bibinfo {year} {2022})}\BibitemShut {NoStop}%
\bibitem [{\citenamefont {Murashita}(2022)}]{murashita2022review}%
  \BibitemOpen
  \bibfield  {author} {\bibinfo {author} {\bibfnamefont {Y.}~\bibnamefont
  {Murashita}},\ }in\ \href {https://doi.org/10.1007/978-981-16-8638-2_2}
  {\emph {\bibinfo {booktitle} {Fluctuation Theorems under Divergent Entropy
  Production and their Applications for Fundamental Problems in Statistical
  Physics}}}\ (\bibinfo  {publisher} {Springer},\ \bibinfo {year} {2022})\ pp.\
  \bibinfo {pages} {7--25}\BibitemShut {NoStop}%
\bibitem [{\citenamefont {Salazar}(2022)}]{salazar2022thermodynamic}%
  \BibitemOpen
  \bibfield  {author} {\bibinfo {author} {\bibfnamefont {D.~S.~P.}\
  \bibnamefont {Salazar}},\ }\href
  {https://doi.org/10.1103/PhysRevE.106.L042101} {\bibfield  {journal}
  {\bibinfo  {journal} {Physical Review E}\ }\textbf {\bibinfo {volume}
  {106}},\ \bibinfo {pages} {L042101} (\bibinfo {year} {2022})}\BibitemShut
  {NoStop}%
\bibitem [{\citenamefont {Jarzynski}(1997)}]{jarzynski1997}%
  \BibitemOpen
  \bibfield  {author} {\bibinfo {author} {\bibfnamefont {C.}~\bibnamefont
  {Jarzynski}},\ }\href {https://doi.org/10.1103/PhysRevLett.78.2690}
  {\bibfield  {journal} {\bibinfo  {journal} {Phys. Rev. Lett.}\ }\textbf
  {\bibinfo {volume} {78}},\ \bibinfo {pages} {2690} (\bibinfo {year}
  {1997})}\BibitemShut {NoStop}%
\bibitem [{\citenamefont {Acconcia}\ and\ \citenamefont
  {Bonan{\c{c}}a}(2015)}]{acconcia2015}%
  \BibitemOpen
  \bibfield  {author} {\bibinfo {author} {\bibfnamefont {T.~V.}\ \bibnamefont
  {Acconcia}}\ and\ \bibinfo {author} {\bibfnamefont {M.~V.~S.}\ \bibnamefont
  {Bonan{\c{c}}a}},\ }\href {https://doi.org/10.1103/PhysRevE.91.042141}
  {\bibfield  {journal} {\bibinfo  {journal} {Phys. Rev. E}\ }\textbf {\bibinfo
  {volume} {91}},\ \bibinfo {pages} {042141} (\bibinfo {year}
  {2015})}\BibitemShut {NoStop}%
\bibitem [{\citenamefont {Bonan{\c{c}}a}(2016)}]{bonancca2016non}%
  \BibitemOpen
  \bibfield  {author} {\bibinfo {author} {\bibfnamefont {M.~V.~S.}\
  \bibnamefont {Bonan{\c{c}}a}},\ }\href
  {https://link.springer.com/article/10.1007/s13538-015-0370-7} {\bibfield
  {journal} {\bibinfo  {journal} {Brazilian journal of physics}\ }\textbf
  {\bibinfo {volume} {46}},\ \bibinfo {pages} {248} (\bibinfo {year}
  {2016})}\BibitemShut {NoStop}%
\bibitem [{\citenamefont {Bonan{\c{c}}a}(2019)}]{bonancca2019approaching}%
  \BibitemOpen
  \bibfield  {author} {\bibinfo {author} {\bibfnamefont {M.~V.~S.}\
  \bibnamefont {Bonan{\c{c}}a}},\ }\href
  {https://iopscience.iop.org/article/10.1088/1742-5468/ab4e92/meta} {\bibfield
   {journal} {\bibinfo  {journal} {Journal of Statistical Mechanics: Theory and
  Experiment}\ }\textbf {\bibinfo {volume} {2019}},\ \bibinfo {pages} {123203}
  (\bibinfo {year} {2019})}\BibitemShut {NoStop}%
\bibitem [{\citenamefont {Jarzynski}(2020)}]{jarzynski2020}%
  \BibitemOpen
  \bibfield  {author} {\bibinfo {author} {\bibfnamefont {C.}~\bibnamefont
  {Jarzynski}},\ }\href {https://doi.org/10.1016/j.physa.2019.122077}
  {\bibfield  {journal} {\bibinfo  {journal} {Physica A: Statistical Mechanics
  and its Applications}\ }\textbf {\bibinfo {volume} {552}},\ \bibinfo {pages}
  {122077} (\bibinfo {year} {2020})}\BibitemShut {NoStop}%
\bibitem [{\citenamefont {Naz{\'e}}\ and\ \citenamefont
  {Bonan{\c{c}}a}(2022)}]{naze2022series}%
  \BibitemOpen
  \bibfield  {author} {\bibinfo {author} {\bibfnamefont {P.}~\bibnamefont
  {Naz{\'e}}}\ and\ \bibinfo {author} {\bibfnamefont {M.~V.~S.}\ \bibnamefont
  {Bonan{\c{c}}a}},\ }\href
  {https://link.springer.com/article/10.1007/s10955-021-02869-0} {\bibfield
  {journal} {\bibinfo  {journal} {Journal of Statistical Physics}\ }\textbf
  {\bibinfo {volume} {186}},\ \bibinfo {pages} {23} (\bibinfo {year}
  {2022})}\BibitemShut {NoStop}%
\bibitem [{\citenamefont {Torrontegui}\ \emph {et~al.}(2013)\citenamefont
  {Torrontegui}, \citenamefont {Ib{\'a}{\~n}ez}, \citenamefont
  {Mart{\'\i}nez-Garaot}, \citenamefont {Modugno}, \citenamefont {del Campo},
  \citenamefont {Gu{\'e}ry-Odelin}, \citenamefont {Ruschhaupt}, \citenamefont
  {Chen},\ and\ \citenamefont {Muga}}]{torrontegui2013shortcuts}%
  \BibitemOpen
  \bibfield  {author} {\bibinfo {author} {\bibfnamefont {E.}~\bibnamefont
  {Torrontegui}}, \bibinfo {author} {\bibfnamefont {S.}~\bibnamefont
  {Ib{\'a}{\~n}ez}}, \bibinfo {author} {\bibfnamefont {S.}~\bibnamefont
  {Mart{\'\i}nez-Garaot}}, \bibinfo {author} {\bibfnamefont {M.}~\bibnamefont
  {Modugno}}, \bibinfo {author} {\bibfnamefont {A.}~\bibnamefont {del Campo}},
  \bibinfo {author} {\bibfnamefont {D.}~\bibnamefont {Gu{\'e}ry-Odelin}},
  \bibinfo {author} {\bibfnamefont {A.}~\bibnamefont {Ruschhaupt}}, \bibinfo
  {author} {\bibfnamefont {X.}~\bibnamefont {Chen}},\ and\ \bibinfo {author}
  {\bibfnamefont {J.~G.}\ \bibnamefont {Muga}},\ }in\ \href
  {https://www.sciencedirect.com/science/article/abs/pii/B9780124080904000025}
  {\emph {\bibinfo {booktitle} {Advances in atomic, molecular, and optical
  physics}}},\ Vol.~\bibinfo {volume} {62}\ (\bibinfo  {publisher} {Elsevier},\
  \bibinfo {year} {2013})\ pp.\ \bibinfo {pages} {117--169}\BibitemShut
  {NoStop}%
\bibitem [{\citenamefont {Gu{\'e}ry-Odelin}\ \emph {et~al.}(2019)\citenamefont
  {Gu{\'e}ry-Odelin}, \citenamefont {Ruschhaupt}, \citenamefont {Kiely},
  \citenamefont {Torrontegui}, \citenamefont {Mart{\'\i}nez-Garaot},\ and\
  \citenamefont {Muga}}]{guery2019shortcuts}%
  \BibitemOpen
  \bibfield  {author} {\bibinfo {author} {\bibfnamefont {D.}~\bibnamefont
  {Gu{\'e}ry-Odelin}}, \bibinfo {author} {\bibfnamefont {A.}~\bibnamefont
  {Ruschhaupt}}, \bibinfo {author} {\bibfnamefont {A.}~\bibnamefont {Kiely}},
  \bibinfo {author} {\bibfnamefont {E.}~\bibnamefont {Torrontegui}}, \bibinfo
  {author} {\bibfnamefont {S.}~\bibnamefont {Mart{\'\i}nez-Garaot}},\ and\
  \bibinfo {author} {\bibfnamefont {J.~G.}\ \bibnamefont {Muga}},\ }\href
  {https://journals.aps.org/rmp/abstract/10.1103/RevModPhys.91.045001}
  {\bibfield  {journal} {\bibinfo  {journal} {Reviews of Modern Physics}\
  }\textbf {\bibinfo {volume} {91}},\ \bibinfo {pages} {045001} (\bibinfo
  {year} {2019})}\BibitemShut {NoStop}%
\bibitem [{\citenamefont {Naz\'e}(2023)}]{naze2023adiabatic}%
  \BibitemOpen
  \bibfield  {author} {\bibinfo {author} {\bibfnamefont {P.}~\bibnamefont
  {Naz\'e}},\ }\href {https://doi.org/10.1103/PhysRevE.107.064114} {\bibfield
  {journal} {\bibinfo  {journal} {Phys. Rev. E}\ }\textbf {\bibinfo {volume}
  {107}},\ \bibinfo {pages} {064114} (\bibinfo {year} {2023})}\BibitemShut
  {NoStop}%
\bibitem [{\citenamefont {Naz{\'e}}(2023)}]{naze2023quantum}%
  \BibitemOpen
  \bibfield  {author} {\bibinfo {author} {\bibfnamefont {P.}~\bibnamefont
  {Naz{\'e}}},\ }\href
  {https://iopscience.iop.org/article/10.1088/1742-5468/ad082e/meta?casa_token=XbITxO6GZ-UAAAAA:0DNN7qA9Ch48KKTR4Iw7DyQIklgA_bBcFzor5dG4v1Capvql0THJ2xLMd_WyRlMdyQzv7C9Xq78hlXTDc_OTLhZ3MLMP}
  {\bibfield  {journal} {\bibinfo  {journal} {Journal of Statistical Mechanics:
  Theory and Experiment}\ }\textbf {\bibinfo {volume} {2023}},\ \bibinfo
  {pages} {113101} (\bibinfo {year} {2023})}\BibitemShut {NoStop}%
\bibitem [{\citenamefont {Naz{\'e}}\ and\ \citenamefont
  {Bonan{\c{c}}a}(2020)}]{naze2020compatibility}%
  \BibitemOpen
  \bibfield  {author} {\bibinfo {author} {\bibfnamefont {P.}~\bibnamefont
  {Naz{\'e}}}\ and\ \bibinfo {author} {\bibfnamefont {M.~V.~S.}\ \bibnamefont
  {Bonan{\c{c}}a}},\ }\href {https://doi.org/10.1088/1742-5468/ab54ba}
  {\bibfield  {journal} {\bibinfo  {journal} {Journal of Statistical Mechanics:
  Theory and Experiment}\ }\textbf {\bibinfo {volume} {2020}},\ \bibinfo
  {pages} {013206} (\bibinfo {year} {2020})}\BibitemShut {NoStop}%
\bibitem [{\citenamefont {Fasano}\ and\ \citenamefont
  {Marmi}(2006)}]{fasano2006analytical}%
  \BibitemOpen
  \bibfield  {author} {\bibinfo {author} {\bibfnamefont {A.}~\bibnamefont
  {Fasano}}\ and\ \bibinfo {author} {\bibfnamefont {S.}~\bibnamefont {Marmi}},\
  }\href
  {https://books.google.com.br/books?hl=en&lr=&id=zDslMYMEGvoC&oi=fnd&pg=PR7&dq=fasano+mechanics&ots=bYC-NoSFzs&sig=nGOTI8NTfTcoGidzw3Qc-79sgBs&redir_esc=y#v=onepage&q=fasano%20mechanics&f=false}
  {\emph {\bibinfo {title} {Analytical mechanics: an introduction}}}\ (\bibinfo
   {publisher} {Oxford University Press},\ \bibinfo {year} {2006})\BibitemShut
  {NoStop}%
\bibitem [{\citenamefont {Jarzynski}(2006)}]{jarzynski2006}%
  \BibitemOpen
  \bibfield  {author} {\bibinfo {author} {\bibfnamefont {C.}~\bibnamefont
  {Jarzynski}},\ }\href {https://doi.org/10.1103/PhysRevE.73.046105} {\bibfield
   {journal} {\bibinfo  {journal} {Phys. Rev. E}\ }\textbf {\bibinfo {volume}
  {73}},\ \bibinfo {pages} {046105} (\bibinfo {year} {2006})}\BibitemShut
  {NoStop}%
\bibitem [{\citenamefont {Acconcia}\ and\ \citenamefont
  {Bonan{\c{c}}a}(2017)}]{acconcia2017microcanonical}%
  \BibitemOpen
  \bibfield  {author} {\bibinfo {author} {\bibfnamefont {T.~V.}\ \bibnamefont
  {Acconcia}}\ and\ \bibinfo {author} {\bibfnamefont {M.~V.~S.}\ \bibnamefont
  {Bonan{\c{c}}a}},\ }\href {https://doi.org/10.1103/PhysRevE.96.062117}
  {\bibfield  {journal} {\bibinfo  {journal} {Physical Review E}\ }\textbf
  {\bibinfo {volume} {96}},\ \bibinfo {pages} {062117} (\bibinfo {year}
  {2017})}\BibitemShut {NoStop}%
\end{thebibliography}%

\end{document}